\documentclass[aps,pra,twocolumn,superscriptaddress,nobalancelastpage]{revtex4-1}
\usepackage{amsmath,amssymb,mathtools}
\usepackage[dvipdfmx]{graphicx}
\usepackage[dvipdfmx,colorlinks=true,bookmarks=true,citecolor=blue,linkcolor=blue,urlcolor=blue,breaklinks=true]{hyperref}

\graphicspath{{fig/}}
\usepackage{dcolumn}
\usepackage{bm}
\usepackage[T1]{fontenc}
\usepackage{textcomp}
\usepackage{color}

\begin{document}

\title{Optomagnonic Josephson effect in antiferromagnets}

\author{Kouki Nakata}
\affiliation{Advanced Science Research Center, Japan Atomic Energy Agency, Tokai,
Ibaraki 319-1195, Japan}

\date{\today}

\begin{abstract}
Combining advanced technologies of optics and antiferromagnetic spintronics, we present a method to realize ultrafast spin transport. The optical Barnett effect provokes quasiequilibrium Bose-Einstein condensates (BECs) of magnons associated with the fully spin-polarized state in insulating antiferromagnets (AFs). This optomagnonic Barnett effect enables us to exploit coherent magnons of high frequency over the conventional ones of (sub-) terahertz associated with the N\'eel magnetic order. We show that the macroscopic coherence of those optical magnon BECs induces a spin current across the junction interface of weakly coupled two insulating AFs, and this optomagnonic Josephson effect realizes ultrafast spin transport. The period of the optomagnonic Josephson oscillation is much shorter than the conventional one of the order of picoseconds. Thus we propose a way to realize ultrafast spin transport in AFs by means of the macroscopic coherence of optical magnon BECs.
\end{abstract}

\maketitle

\section{Introduction}
\label{sec:intro}
For the realization of rapid and efficient transmission of information over electronics,
inventing methods to handle a fast and flexible manipulation of 
spin transport
is a central task in the field of spintronics~\cite{MagnonSpintronics,ReviewMagnon,ReviewOkaKitamura,katsura2,Ultrafast_ST,IshizukaSatoPRL,IshizukaSatoPRB}.
For this goal,
antiferromagnets (AFs)~\cite{AFspintronicsReview,AFspintronicsReview2,AFreviewYT,AFopticsReview,AFpumpingTHz,THzMagnon2,AFsp,AFoscillationTHz,AFoscillationTHz2}  
have an advantage over 
ferromagnets (FMs)~\cite{demokritov,RShindou,RShindou2,Troncoso}
in that spin dynamics is much faster. 
The energy scale of FMs is characterized 
by the macroscopic and classical magnetic dipole interaction
in gigahertz (GHz) regime~\cite{demokritov},
and hence the spin Josephson oscillation~\footnote{
The analogy of the magnetically ordered state to the superconducting state, 
i.e., Josephson effect~\cite{Josephson}, 
was explained in Ref.~\cite{katsura2}.} 
operates of the order of nanoseconds (ns)~\cite{KKPD,katsura2}.
On the other hand,
the energy scale of AFs arises from 
microscopic and quantum-mechanical spin exchange interactions.
Therefore AFs can operate at much higher frequency.
Thus AFs are expected to be the best platform 
for ultrafast transport of spin information~\cite{AFspintronicsReview,AFspintronicsReview2,AFreviewYT,ReviewOkaKitamura,Ultrafast_ST}.
The observation of spin currents 
by means of sub-terahertz (sub-THz) spin pumping in AFs
was reported in Refs.~\cite{THzMagnon2,AFsp}.
Making use of the property of AFs,
spin Josephson effects of THz
associated with the N\'eel magnetic order~\footnote{
See Ref.~\cite{KSJD} for thermomagnetic transport of noncoherent magnons 
associated with the N\'eel magnetic order in the bulk of insulating AFs.}
were theoretically proposed in Refs.~\cite{AFJosephsonTHz,AFJosephsonTHz2}.
The spin Josephson oscillation 
operates of the order of picoseconds (ps).

Another significant development
in the manipulation of magnetism is the utilization of
laser-matter coupling~\cite{Mukai2014APL,LaserPhotoExp,Ciappina2017RepProgPhys,LaserPhotoExp3}.
By means of the optical method~\cite{Kirilyuk,KimelMagnetizationReversalExp4,KimelNatureIFaraday,Kimel7,Kimel8,OtherOpticalBarnett2,OtherOpticalBarnett3}, 
the reversal of magnetization was achieved 
experimentally~\cite{OtherOpticalBarnett,KimelMagnetizationReversalExp2,KimelMagnetizationReversalExp3,KimelMagnetizationReversalExp5},
and an optical analog of the conventional Barnett effect~\cite{Barnett,Barnett2,ReviewSpinMechatronics},
i.e., laser-induced magnetization~\cite{FloquetST2,FloquetST},
was proposed theoretically~\cite{OtherOpticalBarnett2,OtherOpticalBarnett3}.
This optical Barnett effect even provokes
quasiequilibrium Bose-Einstein condensates (BECs) of magnons~\footnote{
See Refs.~\cite{oshikawa,Giamarchi2008NatPhys,bunkov,TotsukaBEC,TotsukaBEC2} for magnon BEC.},
i.e., optical magnon BECs,
and this behavior is especially called
the optomagnonic Barnett effect~\cite{KNST_OBarnett}.
Thus the interdisciplinary field between
optics and magnonics~\cite{OptomagnonicsCavity,OptomagnonicsCavity2,OptomagnonicsCavity3,AFopticsReview},
dubbed optomagnonics,
has been attracting much attention.

\begin{figure}[t]
\begin{center}
\includegraphics[width=6.2cm,clip]{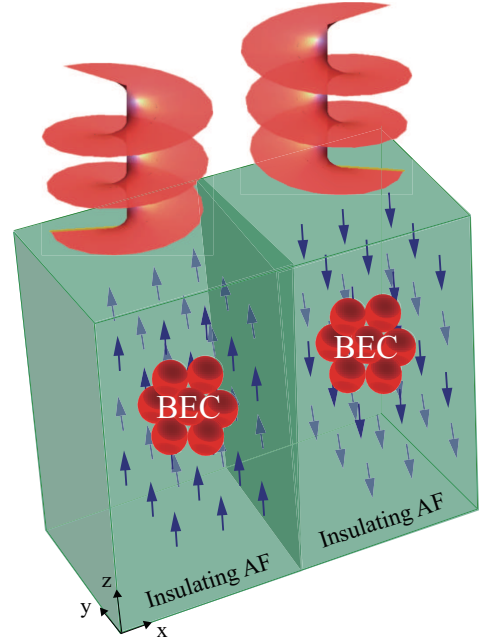}
\caption{Schematic picture of 
the optomagnonic Josephson junction.
The two insulating AFs are separated by a thin film of a nonmagnetic insulator
and weakly exchange-coupled.
We assume an identical material for each AF
subjected to a circularly polarized laser 
with the opposite polarization $\eta =\pm $.
In the vicinity of $\Omega = {\Omega}_{\mathrm{BEC}}$,
the optical Barnett effect realizes the quasiequilibrium magnon BEC,
i.e., the optical magnon BEC,
associated with the fully spin-polarized state
in the high frequency regime,
where spins in the left (right) AF are 
along the +(-) $z$ axis
due to the opposite circular polarization 
(cf., Tables~\ref{tab:Compare} and ~\ref{tab:CompareJosephson0}).
}
\label{fig:junction}
\end{center}
\end{figure}

In this paper using the macroscopic coherence 
of the optical magnon BECs,
we propose a method for the realization of
ultrafast spin transport in insulating AFs.
The optical Barnett effect
realizes the fully spin-polarized state of insulating AFs.
This enables us to exploit coherent magnons 
of high frequency over the conventional ones
of (sub-) THz
associated with the N\'eel magnetic order.
We show that 
the macroscopic coherence of the optical magnon BECs
induces a spin current across the junction interface
of weakly coupled two insulating AFs (Fig.~\ref{fig:junction}).
We refer to this phenomenon 
as the optomagnonic Josephson effect.
The period of the optomagnonic Josephson oscillation
is much shorter 
than the conventional one of the order of picoseconds.
This ultrafast phenomenon intrinsic to AFs, 
the optomagnonic Josephson effect,
is the result from the confluence of 
optics and antiferromagnetic magnonics.
We also discuss an experimental scheme for the observation.

We remark that 
in this paper using the scheme of Refs.~\cite{FloquetST2,FloquetST},
we consider transport of the optical magnon BECs in AFs,
i.e., magnon BECs out of equilibrium,
associated with the fully spin-polarized state
of high frequency 
over the conventional one
of (sub-) THz associated with the N\'eel magnetic order~\cite{Supple}
(cf., Sec.~\ref{subsec:Sec.3}).
See Refs.~\cite{oshikawa,bunkov} for magnon BECs in equilibrium
subjected to a static magnetic field.

This paper is organized as follows. 
In Sec.~\ref{sec:Sec.2}
we quickly review the optical Barnett effect.
Then we investigate the prominent application, 
the optomagnonic Josephson effect, 
in Sec.~\ref{sec:Sec.2-2}
and give an estimate for the experimental feasibility
in Sec.~\ref{sec:Sec.9}.
Finally, we remark on several issues in Sec.~\ref{sec:Sec.7}
and summarize in Sec.~\ref{sec:Sec.11}.
Technical details are described in the Appendices.

\section{Optical Barnett effect}
\label{sec:Sec.2}
Before going to the main subject, for readers' convenience
let us quickly review the mechanism of the laser-induced magnetization~\cite{FloquetST2,FloquetST},
i.e., the optical Barnett effect~\cite{OtherOpticalBarnett2,OtherOpticalBarnett3}.
See Refs.~\cite{FloquetST2,FloquetST} for details~\footnote{
See also Ref.~\cite{KNST_OBarnett}
for the difference from the inverse Faraday effect~\cite{Kirilyuk,KimelNatureIFaraday}.
},
especially for the importance of 
modulating laser frequency adiabatically
by the chirping technique~\cite{chirping,chirping2}.

We consider a magnetic insulator with a large electronic gap
described by the Hamiltonian 
$\mathcal{H}_{0}$ which has the $U(1)$ symmetry 
about an axis,
and
we take it the $z$ axis for convenience. 
Due to the large electronic gap,
spins in the circularly polarized laser
interact only with the magnetic component of the laser
through the Zeeman coupling.
We take the polarization plane of the laser 
as the $xy$ plane.
We adiabatically apply the laser of 
the frequency $\Omega>0$
with the magnetic field amplitude $B_{0}>0$.
For the generation of the optical Barnett effect,
the driving field amplitude $B_0>0$
should take a nonzero value
$ B_0 \neq 0 $
of
being strong enough that 
$ B_0 > |u_0|  $,
where $u_0$ is the potential energy of magnons in the lattice
formed by surroundings 
(e.g., phonons and impurities, etc.).
Since throughout this paper we assume the clean magnet at low temperatures,
the condition is satisfied.
The spin system subjected to the laser
is described by the time-periodic Hamiltonian, 
$ {\hat{\mathcal{H}}}(t)={\hat{\mathcal{H}}}_{0}
   -B_{0}[{\hat{S}}_{\mathrm{tot}}^{x}\cos(\Omega t)
     +\eta {\hat{S}}_{\mathrm{tot}}^{y}\sin(\Omega t)]$,
where 
the sign $\eta=+(-)$ represents the left (right) circular polarization
and 
$S_{\mathrm{tot}}^{x(y,z)} := \sum_{j}S_{j}^{x(y,z)}$
is the summation over spin operators on all the spin sites. 
Using the unitary transformation,
we obtain an effective static Hamiltonian 
in the rotational frame~\cite{RMPcorotating}
of the frequency $\eta \Omega$ around the $z$ axis as~\cite{FloquetST2,FloquetST}
\begin{align}
 {\hat{{\mathcal{H}}}}_{\mathrm{eff}}
   = {\hat{{\mathcal{H}}}}_{0}
   -\eta \hbar \Omega  {\hat{S}}_{\mathrm{tot}}^{z}
   +O(B_0).
   \label{eq:Hamil_eff}
\end{align}
Hereafter we assume a weak laser field 
$B_{0}\ll \hbar\Omega$
where the $B_{0}S_{\mathrm{tot}}^{x}$ term is negligibly small.
In Eq.~\eqref{eq:Hamil_eff}, the effective magnetic field $\Omega/\gamma $
with the gyromagnetic ratio $\gamma $ may be regarded as an optical analog~\cite{OtherOpticalBarnett2,OtherOpticalBarnett3}
of the conventional Barnett field~\cite{Barnett,Barnett2,ReviewSpinMechatronics}
along the $z$ axis.
This optical Barnett field develops 
the total magnetization of magnets.
The direction of the optical Barnett field is controllable 
by means of the change of the laser chirality $\eta =\pm $.
The effective static Hamiltonian
${\hat{{\mathcal{H}}}}_{\mathrm{eff}}$ [Eq.~\eqref{eq:Hamil_eff}]
has the $U(1)$ symmetry.

We remark that
modulating laser frequency 
$\Omega$ slowly enough
through the chirping technique~\cite{chirping,chirping2},
the adiabatic time-evolution
is realized~\cite{FloquetST2,FloquetST}.
Therefore the spin configuration is determined in the way that
the energy of the Hamiltonian ${\hat{{\mathcal{H}}}}_{\mathrm{eff}}$ [Eq.~\eqref{eq:Hamil_eff}]
per a site is minimized.

\section{Optomagnonic Josephson effect}
\label{sec:Sec.2-2}

\subsection{Optical magnon BEC in AF}
\label{subsec:Sec.3}
We apply the optical Barnett effect [Eq.~\eqref{eq:Hamil_eff}] to an insulating AF
described by the Hamiltonian,
\begin{align}
 {\hat{{\mathcal{H}}}}_{0}
   =J\sum_{\langle i, j \rangle}
     \boldsymbol{{\hat{S}}}_{i} \cdot
     \boldsymbol{{\hat{S}}}_{j}
     +D\sum_{i}
       ({\hat{S}}_i^{z})^{2},
       \label{eq:Hamil_0}
\end{align}
where 
$\boldsymbol{{\hat{S}}}_{i(j)}=({\hat{S}}_{i(j)}^{x}, {\hat{S}}_{i(j)}^{y}, {\hat{S}}_{i(j)}^{z})$ 
represents the spin operator on the $i(j)$-th site 
having the spin quantum number $S$,
$J>0$ is the exchange interaction between the nearest neighbor spins 
$\langle i, j \rangle$,
and $D>0$ is the easy-plane single ion anisotropy 
that stabilizes the N\'eel magnetic order on the $xy$ plane. 
Hereafter, we consider a cubic lattice.

First, 
the optical Barnett field [Eq.~\eqref{eq:Hamil_eff}] develops the total magnetization of the AF
along the $z$ axis continuously.
We find from a microscopic calculation~\cite{Supple} that
in the high frequency regime $\Omega  >\Omega _{\mathrm{BEC}}$ defined as
\begin{align}
 \Omega _{\mathrm{BEC}}:=2(6J+D)S/\hbar ,
\label{eq:BEC}
\end{align}
spins are fully polarized along the $\eta $z axis,
and confirmed the absence of the first order transition
in the vicinity of ${\Omega}_{\mathrm{BEC}}$~\cite{Supple}.
This assures the validity of 
the description in terms of the magnon picture~\cite{HP}. 
Hence we move to the analysis by the magnon theory next.
For the details of the calculation, see the Appendices~\cite{Supple}.

Note that magnons acquire
the effective magnetic field in the corotating frame as~\cite{Supple}
${\mathbf{B}} _{\mathrm{eff}}
=(B _{\mathrm{eff}}^x, 0, B _{\mathrm{eff}}^z) $,
where
$B _{\mathrm{eff}}^x:= B_0$
along the $x$ axis
and
$B _{\mathrm{eff}}^z:= \hbar (\Omega -\Omega _{\mathrm{BEC}})+6JS$
along the $z$ axis.
Since
$6JS  \gg B_0  $ in general,
the effective magnetic field along the $x$ axis
$B _{\mathrm{eff}}^x= B_0$
is negligibly small
compared with the $z$ component
$B _{\mathrm{eff}}^x \ll B _{\mathrm{eff}}^z  $
even in the vicinity of
$\Omega \approx \Omega _{\mathrm{BEC}} $.
Throughout this paper we work under the assumption that
$ 0< |u_0| < B_0 \ll  \hbar \Omega, 6JS$.

Next,
decreasing the frequency $\Omega$ 
from above the critical value $\Omega _{\mathrm{BEC}}$ 
in which spins are full polarized, 
magnon BEC transition is provoked on the point
$\Omega =\Omega _{\mathrm{BEC}}$
and magnons of $\pi$ mode,
${\boldsymbol{k}}=\boldsymbol{\pi}:=
(\pi/a,\pi/a, \pi/a)$,
begin to condensate,
where
${\boldsymbol{k}}$ is the wavenumber
and
$a$ is the lattice constant.
We refer to this behavior
as the optomagnonic Barnett effect,
and
the resulting magnon condensate 
as the optical magnon BEC~\cite{KNST_OBarnett}.
In the frequency $\Omega <\Omega _{\mathrm{BEC}}$,
the AF acquires a transverse component of local magnetization
associated with the spontaneous $U(1)$ symmetry breaking,
and thus forms a macroscopic coherent state.
The optical magnon BEC state is described 
by the effective Hamiltonian
in the rotational frame 
of the frequency $\eta \Omega$ around the $z$ axis as~\cite{Supple}
\begin{align}
{\hat{{\mathcal{H}}}}_{\mathrm{eff}}(\boldsymbol{k}=\boldsymbol{\pi})
   =\hbar (\Omega -\Omega _{\mathrm{BEC}}){\hat{a}}^{\dagger }_{\boldsymbol{\pi}}{\hat{a}}_{\boldsymbol{\pi}} 
   +U{\hat{a}}^{\dagger }_{\boldsymbol{\pi}}{\hat{a}}^{\dagger }_{\boldsymbol{\pi}}{\hat{a}}_{\boldsymbol{\pi}}{\hat{a}}_{\boldsymbol{\pi}},
\label{eq:Hamil_BECa}
\end{align}
where
${\hat{a}}^{(\dagger) }_{\boldsymbol{\pi}}$ 
is the bosonic annihilation (creation) operator for
magnons of the $\pi$ mode in condensation,
\begin{align}
U:=\frac{\hbar \Omega _{\mathrm{BEC}}}{2S{{N}}}
\label{eq:UBEC}
\end{align}
represents the magnitude of magnon-magnon interactions,
and ${N}$ is the number of spin sites. 
The magnon-magnon interaction is repulsive $U>0$.
Therefore the magnon BEC characterized by
the expectation value
$\langle {\hat{a}}_{\boldsymbol{\pi}}  \rangle\not=0$
are stable~\cite{oshikawa}.

Finally,
the effective Hamiltonian for the optical magnon BEC 
in the rotational frame
is recast into the Hamiltonian
in the original stationary reference frame as~\cite{Supple}
 \begin{align}
 {\hat{{\mathcal{H}}}}_{\boldsymbol{k}=\boldsymbol{\pi}}
   =-\hbar \Omega _{\mathrm{BEC}} {\hat{b}}^{\dagger }_{\boldsymbol{\pi}}{\hat{b}}_{\boldsymbol{\pi}} 
   +U{\hat{b}}^{\dagger }_{\boldsymbol{\pi}}{\hat{b}}^{\dagger }_{\boldsymbol{\pi}}{\hat{b}}_{\boldsymbol{\pi}}{\hat{b}}_{\boldsymbol{\pi}},
   \label{eq:Hamil_BEC2}
\end{align}
where
$\hat{b}_{\boldsymbol{\pi}}^{(\dagger)}$ is the magnon operator in the reference frame 
and
$\hat{a}_{\boldsymbol{k}}^{(\dagger)}
 =\hat{R}^{\dagger}\hat{b}_{\boldsymbol{k}}^{(\dagger)}\hat{R}$
for
$\hat{R}:=\exp(\eta i\Omega t\hat{S}_{\mathrm{tot}}^{z})$.
This Hamiltonian depends solely on the material parameters $\Omega _{\mathrm{BEC}}  $ 
[Eqs.~\eqref{eq:BEC} and~\eqref{eq:UBEC}],
while it is independent of laser frequency.
Note that
the number of magnon BECs
is characterized as a function of laser frequency~\cite{Supple}.

We remark that 
condensation of the $\pi$ mode magnons
does not induce a Josephson-like effect in the single AF
since the $xy$ components of the nearest neighbor spins
are in the opposite direction
and this results in ${\mathrm{sin}}(\pm  \pi)=0$.

\subsection{Optomagnonic Josephson junction}
\label{subsec:Sec.4}
In this paper 
using the optical magnon BEC
in the vicinity of 
$ {\Omega } = {\Omega }  _{\mathrm{BEC}} $,
we investigate the application
of the macroscopic coherence 
to ultrafast spin transport.
To this end, 
we consider a junction of weakly exchange-coupled two insulating AFs shown in Fig.~\ref{fig:junction}.
The two AFs are separated by a thin film of a nonmagnetic insulator~\cite{AFJosephsonTHz}
and weakly exchange-coupled.
The AFs are subjected to a circularly polarized laser 
of the frequency $\Omega <\Omega _{\mathrm{BEC}}$
with the opposite polarization $\eta =\pm $.
Thus optical magnon BECs are realized and
spins of the left (right) AF are aligned along the +(-) $z$ axis
due to the opposite circular polarization (Table~\ref{tab:Compare}).

First, 
we assume an identical material for each AF.
From Eq.~\eqref{eq:Hamil_BEC2}
the optical magnon BEC in the left (right) AF
is described by the Hamiltonian ${\hat{\mathcal{H}}}_{\mathrm{L(R)}}$
in the original stationary frame as~\cite{Supple}
\begin{subequations}
\begin{align}
{\hat{\mathcal{H}}}_{\mathrm{L}}
    =&\hbar \Omega_{\mathrm{L}} {\hat{b}}^{\dagger }_{\mathrm{L}} {\hat{b}}_{\mathrm{L}}
    +U_{\mathrm{L}} {\hat{b}}^{\dagger }_{\mathrm{L}} {\hat{b}}^{\dagger }_{\mathrm{L}} {\hat{b}}_{\mathrm{L}}{\hat{b}}_{\mathrm{L}},  \\
{\hat{\mathcal{H}}}_{\mathrm{R}}
    =&\hbar \Omega_{\mathrm{R}} {\hat{b}}^{\dagger }_{\mathrm{R}} {\hat{b}}_{\mathrm{R}}
    +U_{\mathrm{R}} {\hat{b}}^{\dagger }_{\mathrm{R}} {\hat{b}}^{\dagger }_{\mathrm{R}} {\hat{b}}_{\mathrm{R}}{\hat{b}}_{\mathrm{R}}, 
\label{eq:two-state}
\end{align}
\end{subequations}
where
\begin{subequations}
\begin{align}
\Omega_{\mathrm{L}}  =& \Omega_{\mathrm{R}} := - \Omega_{\mathrm{BEC}} <0, \label{eq:theta_LR}  \\
U_{\mathrm{L}} =& U_{\mathrm{R}}:=U >0,
\end{align}
\end{subequations}
and
${\hat{b}}^{(\dagger) }_{\mathrm{L(R)}}$ 
is the bosonic annihilation (creation) operator 
for
magnon condensates of the $\pi$ mode
in the left (right) AF.

\begin{table}[t]
\centering
\caption{Comparison of the optical magnon BECs 
in the junction of AFs shown in Fig.~\ref{fig:junction}.
}
\begin{tabular}{l|cc}
\hline
  & Left BEC & Right BEC \\
\hline
Frequency  &  $O(10)$ THz  & $O(10)$ THz \\
Circular polarization  & $\eta =+$ & $\eta =-$  \\
Spin polarization & $+z$ axis & $-z$ axis  \\
Spin angular momentum & $-$ & $+$ \\
Macroscopic coherent state  & $ b_{\mathrm{L}}=\sqrt{N_{\mathrm{L}}} {\mathrm{e}}^{i\theta _{\mathrm{L}}}$  & 
$ b_{\mathrm{R}}=\sqrt{N_{\mathrm{R}}} {\mathrm{e}}^{-i\theta _{\mathrm{R}}}$  \\
\hline
\end{tabular}
\label{tab:Compare}
\end{table}

Next, we focus on the junction interface connecting the two AFs.
Due to a finite overlap of the wave functions of the localized spins that reside on the relevant two-dimensional boundaries of each insulator,
there exists in general a finite exchange interaction 
between the boundary spins~\cite{KKPD,magnonWF}.
This induces a tunneling process of magnons 
across the junction interface.
Let us denote the tunneling amplitude as $\mid K\mid $.
Since two AFs are separated by a thin film of a nonmagnetic insulator~\cite{AFJosephsonTHz},
those are weakly exchange-coupled.
In the tunneling limit, 
the energy scale is assumed to be 
$\mid K\mid \ll   \mid  \hbar \Omega _{\mathrm{L(R)}} \mid  $
and $\mid K\mid \ll   J  $.
During the tunneling process, 
the spin angular momentum is exchanged 
between the left and the right BECs via magnons.
Since spins in the left (right) AF are aligned 
along the +(-) $z$ axis
by the opposite circular polarization $\eta =\pm $,
the magnon of the left BEC carries the spin angular momentum 
$\delta S^z_{\mathrm{L}}=-$,
while that of the right BEC carries 
the opposite
$\delta S^z_{\mathrm{R}}=+$.
Therefore,
within the low energy regime (i.e., in the lowest order of magnon operators),
from the conservation law of the total spin angular momentum
the tunneling process at the junction interface 
is effectively described 
by the Hamiltonian~\cite{Supple}
${\hat{V}}=-K( {\hat{b}}_{\mathrm{L}} {\hat{b}}_{\mathrm{R}}+{\hat{b}}^{\dagger }_{\mathrm{L}}{\hat{b}}^{\dagger }_{\mathrm{R}})$,
where ${\mathrm{sgn}}(K)=\pm $ in general~\cite{AFJosephsonTHz,Bruno}.
Note that the total number of magnons is not conserved 
due to this tunneling process.

Finally,
the total Hamiltonian 
for the optomagnonic Josephson junction,
Fig.~\ref{fig:junction},
is summarized as
${\hat{\mathcal{H}}}_{\mathrm{tot}}={\hat{\mathcal{H}}}_{\mathrm{L}}+{\hat{\mathcal{H}}}_{\mathrm{R}}+{\hat{V}}  $.
See the Appendices
for the tunneling amplitude in spin language~\cite{Supple}.

\subsection{Optomagnonic Josephson equation}
\label{subsec:Sec.5}
Starting from the Hamiltonian 
${\hat{\mathcal{H}}}_{\mathrm{tot}}$ 
of the junction system,
we derive the spin Josephson equation 
of the optomagnonic Barnett effect.
First,
since the optical magnon BEC is a macroscopic coherent state, 
it acquires a macroscopic coherence
$ \langle {\hat{b}}_{\mathrm{L(R)}}(t) \rangle =:b_{\mathrm{L(R)}}(t)\not=0$
characterized as
$b_{\mathrm{L}}(t) =\sqrt{N_{\mathrm{L}}(t)} {\mathrm{e}}^{i\theta _{\mathrm{L}}(t)} \in {\mathbb{C}}$
and 
$b_{\mathrm{R}}(t) =\sqrt{N_{\mathrm{R}}(t)} {\mathrm{e}}^{-i\theta _{\mathrm{R}}(t)} \in {\mathbb{C}}$,
where $N_{\mathrm{L(R)}}(t)$ is the number of magnon BECs 
in the left (right) insulator
and $\theta _{\mathrm{L(R)}}$ is the phase~\cite{katsura2}.
The sign change in the phase, 
$i\theta _{\mathrm{L}}(t)$ and $-i\theta _{\mathrm{R}}(t)$,
arises from the fact that 
spins of the left (right) AF are aligned along the +(-) $z$ axis
by the opposite circular polarization $\eta =\pm $;
the spin raising operation corresponds to the magnon annihilation 
in the left BEC,
while to the magnon creation in the right BEC (Table.~\ref{tab:Compare}).

Next,
using the Heisenberg equation of motion 
for ${\hat{\mathcal{H}}}_{\mathrm{tot}}$
and taking the expectation value 
$ \langle {\hat{b}}_{\mathrm{L(R)}}(t) \rangle =:b_{\mathrm{L(R)}}(t)$,
we derive~\cite{Supple} the two-state model~\cite{smerzi,KKPD} 
for the optical magnon BECs 
\begin{subequations}
\begin{align}
i\hbar \frac{d{{{b}}}_{\mathrm{L}}(t)}{dt}
    =& \hbar \Omega_{\mathrm{L}}  {{b}}_{\mathrm{L}} 
      + 2 U_{\mathrm{L}} N_{\mathrm{L}} {{b}}_{\mathrm{L}} 
       - K {{b}}^{\dagger }_{\mathrm{R}}, 
\label{eq:two-state2} \\
i\hbar \frac{d{{{b}}}_{\mathrm{R}}(t)}{dt}
    =& \hbar \Omega_{\mathrm{R}}  {{b}}_{\mathrm{R}} 
      + 2 U_{\mathrm{R}} N_{\mathrm{R}} {{b}}_{\mathrm{R}} 
       - K {{b}}^{\dagger }_{\mathrm{L}}.
\label{eq:two-state2-2}
\end{align}
\end{subequations}
Then we divide Eqs.~\eqref{eq:two-state2} and~\eqref{eq:two-state2-2}
into the real and imaginary parts as
\begin{subequations}
\begin{align}
\frac{d}{dt}&[N_{\mathrm{L}}(t)-N_{\mathrm{R}}(t)]
   =0,  
\label{eq:two-stateN-} \\
\frac{d}{dt}&[N _{\mathrm{L}}(t)+N _{\mathrm{R}}(t)]
   =-\frac{4K}{\hbar}\sqrt{N_{\mathrm{L}}N_{\mathrm{R}}}{\mathrm{sin}}(\theta _{\mathrm{R}} - \theta _{\mathrm{L}}),
\label{eq:N_{-}}  \\
-\hbar & \frac{d\theta _{\mathrm{L}}(t)}{dt}
    = (\hbar \Omega _{\mathrm{L}}+2 U_{\mathrm{L}} N_{\mathrm{L}}) 
    - K\sqrt{\frac{N_{\mathrm{R}}}{{N_{\mathrm{L}}}}} {\mathrm{cos}}(\theta _{\mathrm{R}} - \theta _{\mathrm{L}}), 
\label{eq:two-state3-2} \\
\hbar & \frac{d\theta _{\mathrm{R}}(t)}{dt}
    = (\hbar \Omega _{\mathrm{R}}+2 U_{\mathrm{R}} N_{\mathrm{R}}) 
    - K\sqrt{\frac{N_{\mathrm{L}}}{{N_{\mathrm{R}}}}} {\mathrm{cos}}(\theta _{\mathrm{R}} - \theta _{\mathrm{L}}).
\label{eq:two-state3}
\end{align}
\end{subequations}
Eqs.~\eqref{eq:two-stateN-} and~\eqref{eq:N_{-}} mean that
the total number of magnons in condensation 
$ N_{+}(t):=N_{\mathrm{L}}(t)+N_{\mathrm{R}}(t) $ is not conserved
due to the tunneling process,
while the total spin angular momentum $ N_{-}:=N_{\mathrm{L}}(t)-N_{\mathrm{R}}(t) $ is conserved.
This ensures that the left BEC acquires the spin angular momentum lost in the right BEC, and vice versa.
The initial condition $N_{+}(0)$ and $N_{-}(0)$,
i.e., $N_{\mathrm{L(R)}}(0)$,
is characterized as a function of laser frequency~\cite{Supple}.

\begin{figure}[t]
\begin{center}
\includegraphics[width=6.5cm,clip]{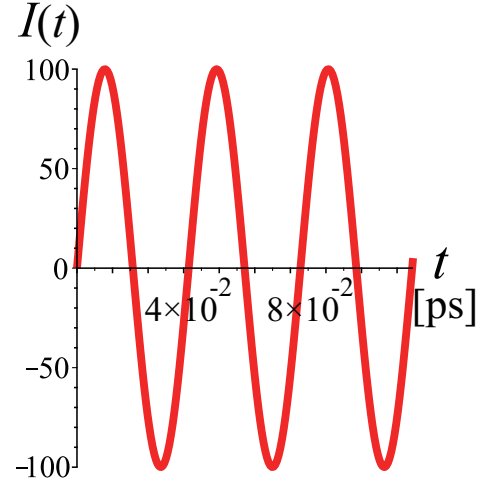}
\caption{
Plot of the rescaled Josephson spin current,
$I(t):=[\hbar /(2K)][dz/(dt)]$,
as a function of time
in the vicinity of 
$\Omega = \Omega _{\mathrm{BEC}} = 75$ THz
obtained by numerically solving 
Eqs.~\eqref{eq:JosephsonEqz}
and~\eqref{eq:JosephsonEq}
with the initial condition
$z(0)=10^2$ and $\theta (0)=0$
for experimental values given in the main text.
Assuming 
$K=0.375$ $\mu$ev,
the period of the optomagnonic Josephson oscillation
becomes $O(10^{-2})$ ps.
}
\label{fig:OMJosephson}
\end{center}
\end{figure}

Finally, we introduce the variable $ z(t)$ 
to describe the spin current across the junction interface as
$ z(t):={N_{+}(t)}/{N_{-}}$,
and define the relative phase as
$ \theta (t) := \theta _{\mathrm{R}}(t)-\theta _{\mathrm{L}}(t)$,
where $\mid z(t)\mid \geq 1$ by definition.
In this work, without loss of generality 
we assume the initial condition $N_{-}(0)>0$ for convenience. 
Since $N_{-}:=N_{\mathrm{L}}-N_{\mathrm{R}}$ is constant,
this ensures $z(t)\geq 1$. 
In terms of the variables $z(t)$ and $\theta(t)$,
Eqs.~\eqref{eq:two-stateN-}
-~\eqref{eq:two-state3}
are summarized  as~\cite{Supple}
\begin{subequations}
\begin{align}
\frac{dz(t)}{dt}
    =& -\frac{2K}{\hbar}\sqrt{z(t)^2-1}{\mathrm{sin}}\theta (t),
\label{eq:JosephsonEqz} \\
\frac{d\theta (t)}{dt}
    =& \Big[(\Omega _{\mathrm{L}}+\Omega _{\mathrm{R}})
      +\frac{U _{\mathrm{L}}-U _{\mathrm{R}}}{\hbar}N_{-} \Big] 
\label{eq:JosephsonEq} \\
    +&\Big(\frac{U _{\mathrm{L}}+U _{\mathrm{R}}}{\hbar}N_{-} \Big) z(t) 
    - \frac{2K}{\hbar}\frac{z(t)}{\sqrt{z(t)^2-1}}{\mathrm{cos}}\theta (t).  \nonumber
\end{align}
\end{subequations}
This is the spin Josephson equation 
for the junction of the AFs
shown in Fig.~\ref{fig:junction}
subjected to the optomagnonic Barnett effect.
We refer to this equation as
the optomagnonic Josephson equation.
Eq.~\eqref{eq:JosephsonEqz}, 
$dz(t)/(dt)\propto {\mathrm{sin}}\theta (t)$, 
describes the Josephson spin current 
across the junction interface,
and
Eq.~\eqref{eq:JosephsonEq},
$d\theta/(dt)$,
shows the time-evolution of the relative phase.
The numerical plot of the Josephson spin current
is depicted in 
Fig.~\ref{fig:OMJosephson}.
Eq.~\eqref{eq:JosephsonEqz}
means that
due to the macroscopic coherence of 
the optical magnon BEC,
the spin current of 
$O(K)$
arises from the phase difference.
This is in contrast to the junction of
noncondensed magnons~\cite{magnonWF},
where
spin currents of 
$O(K^2)$
are generated,
e.g.,
by the temperature difference.
Note that being independent of the sign of the parameter $K$,
the spin Josephson effect is induced;
Eqs.~\eqref{eq:JosephsonEqz} 
and~\eqref{eq:JosephsonEq}
are invariant under the transformation
$K\rightarrow -K $
and
$\theta (t) \rightarrow  \theta (t)+\pi $.
Thus the effect of the sign change, ${\mathrm{sgn}}(K)=\pm $,
is absorbed into the initial condition of the relative phase
and the transport property remains unchanged essentially.

We remark that 
the quasiequilibrium magnon BEC of $\pi$ mode
corresponds to a macroscopic coherent spin precession
where the $xy$ components of the nearest neighbor spins
in the single AF are in the opposite direction. 
This does not affect the Josephson effect 
in the junction of the AFs
since the Josephson equations
[Eqs.~\eqref{eq:JosephsonEqz} and~\eqref{eq:JosephsonEq}]
are invariant under the transformation
$\theta (t) \rightarrow  \theta (t)+2\pi $.

\subsection{Optomagnonic Josephson spin current}
\label{subsec:Sec.6}
The transition point for the optical magnon BEC of the AF
amounts to
$\Omega _{\mathrm{BEC}}=O(10)$ THz.
Under some conditions,
Eq.~\eqref{eq:JosephsonEq} 
approximately
reduces to
$ {d\theta (t)}/({dt})\approx  \Omega _{\mathrm{L}}+\Omega _{\mathrm{R}}$
and essentially results in
$ \theta (t) = ( \Omega _{\mathrm{L}}+\Omega _{\mathrm{R}})t
+\theta (0) $.
From Eq.~\eqref{eq:JosephsonEqz} we find that
${dz(t)}/(dt) \propto {\mathrm{sin}}[( \Omega _{\mathrm{L}}+\Omega _{\mathrm{R}})t+\theta (0)]$.
The period of the optomagnonic Josephson oscillation 
is estimated to be
${2\pi}/\mid {\Omega _{\mathrm{L}}+\Omega _{\mathrm{R}}}\mid =O(10^{-2})$ ps.
Thus ultrafast spin transport is realized in AFs.
We refer to this phenomenon as the optomagnonic Josephson effect.
The analytic estimation agrees with the numerical calculation 
shown in Fig.~\ref{fig:OMJosephson}.

\begin{table}[t]
\centering
\caption{
The comparison of the spin Josephson effect
in the AF-AF junction;
the conventional spin Josephson effect
of Ref.~\cite{AFJosephsonTHz,AFJosephsonTHz2}
and the optomagnonic Josephson effect
of this study
(Fig.~\ref{fig:junction}).
}
\begin{tabular}{l|cc}
\hline
  & Conventional & Optomagnonic \\
\hline
Order & N\'eel magnetic order & Fully spin-polarized state  \\
Coherence & e.g., AF resonance & Optical Barnett effect \\
Frequency & Sub-THz or $O(1)$ THz & $O(10)$ THz \\
Period & $O(1)$ ps  & $ O(10^{-2})$ ps  \\
\hline
\end{tabular}
\label{tab:CompareJosephson0}
\end{table}

\section{Experimental feasibility}
\label{sec:Sec.9}
For an estimate,
we assume the following experiment parameter values
for an insulating AF, NiO, 
as~\cite{ohnuma,NiO_2,NiO_3}
$J=6.3$ meV,
$D=0.1$ meV,
and $S=1$.
We find that 
the transition point for the optical magnon BEC amounts to
$ \Omega _{\mathrm{BEC}}=75$ THz,
which is much higher than the conventional one
$ \Omega _{\mathrm{res}}=O(1)$ THz or sub-THz
for the antiferromagnetic resonance
associated with the N\'eel magnetic order~\cite{THzMagnon2,AFsp,AFJosephsonTHz,AFJosephsonTHz2}.
The numerical plot of
the optomagnonic Josephson effect
is in Fig.~\ref{fig:OMJosephson} 
with the parameter values in its caption.
Given these estimates we expect that, 
while being challenging,
our proposal will be within experimental reach
with current device and measurement technologies,
e.g.,
femtosecond mid-infrared pump-probe spectroscopy~\cite{PumpProbeSpectroscopy,PumpProbeSpectroscopy2Attosecond,Kirilyuk,THzspectroscopy,THzspectroscopy2} 
for the ultrafast spin dynamics,
and Brillouin light scattering (BLS)~\cite{demokritov,MagnonSupercurrent} 
for the optical magnon BEC 
and 
the resulting Josephson effect.
Ref.~\cite{MagnonJosephsonTC}
reported the observation of the ac Josephson effect of magnon BECs
in $^3$He-B.
We expect from a theoretical viewpoint that
to use the inverse spin Hall effect~\cite{ISHE1}
by attaching a metal to the insulating AF
will be one of the most promising strategies
for the observation of the magnon Josephson effect in magnets.

\section{Discussion}
\label{sec:Sec.7}

We remark on the difference
between this study
and
other works on spin Josephson effects
proposed in
Refs.~\cite{AFJosephsonTHz,AFJosephsonTHz2,KKPD}.
In the conventional low frequency region of the AF,
coherent magnons associated with the N\'eel magnetic order
are available, 
e.g., by antiferromagnetic resonance~\cite{AFoscillationTHz}.
However, the frequency of the conventional coherent magnons
can amount only to $ \Omega _{\mathrm{res}}=O(1)$ THz 
or sub-THz~\cite{THzMagnon2,AFsp}.
Thus the period of the resulting Josephson-like effect
is estimated to be 
$O(1)$ ps~\cite{AFJosephsonTHz,AFJosephsonTHz2}.
Note that the quasiequilibrium magnon BEC 
of Ref.~\cite{KKPD}
through microwave pumping in FMs
is in the $O(1)$ GHz regime~\cite{demokritov},
in much lower frequency,
where
the spin Josephson oscillation
operates of the order of ns.

\begin{table}[t]
\centering
\caption{
The comparison of the spin Josephson effect;
the FM-FM junction 
of Ref.~\cite{KKPD}
through microwave pumping
and 
the AF-AF junction 
of this work (Fig.~\ref{fig:junction})
through the optomagnonic Barnett effect,
where
$ {\textrm{sgn}}(\omega _{\mathrm{L}})
= {\textrm{sgn}}(\omega _{\mathrm{R}})$
and
$ {\textrm{sgn}}(\Omega _{\mathrm{L}})
= {\textrm{sgn}}(\Omega _{\mathrm{R}})$,
respectively.
See Ref.~\cite{KKPD} 
for the details of the frequency 
$\omega _{\mathrm{L(R)}}$
of the left (right) BEC in the FM-FM junction.
}
\begin{tabular}{l|cc}
\hline
  & FM-FM junction & AF-AF junction \\
\hline
Magnon BEC & Microwave pumping & Optical Barnett  \\
Total magnon number & Conserved & Nonconserved \\
Spin angular momentum & Conserved & Conserved \\
Josephson oscillation 
& $  {\mathrm{sin}}[(\omega _{\mathrm{L}}-\omega _{\mathrm{R}})t ]$  
& $  {\mathrm{sin}}[(\Omega _{\mathrm{L}}+\Omega _{\mathrm{R}})t]$ \\
Frequency & $O(1)$ GHz & $O(10)$ THz \\
Period & $O(1)$ ns & $O(10^{-2})$ ps \\
\hline
\end{tabular}
\label{tab:CompareJosephson}
\end{table}

There is a distinction also in the spin Josephson equation
due to the direction of the macroscopic coherent spin precession in the BEC phase.
In the same way as in Ref.~\cite{KKPD},
the Josephson spin current of this work
[Eqs.~\eqref{eq:JosephsonEqz} and~\eqref{eq:JosephsonEq}]
is proportional to ${\mathrm{sin}}\theta(t)$.
However, in contrast to Ref.~\cite{KKPD},
the relative phase of this work is described
essentially as the sum of the frequency 
$\Omega _{\mathrm{L(R)}}$
of the left (right) BEC as
$ \theta (t) = \theta _{\mathrm{R}}(t)-\theta _{\mathrm{L}}(t)
=( \Omega _{\mathrm{L}}+\Omega _{\mathrm{R}})t+\theta (0)$
where
$ {\textrm{sgn}}(\Omega _{\mathrm{L}})
= {\textrm{sgn}}(\Omega _{\mathrm{R}})$
[Eq.~\eqref{eq:theta_LR}].
This arises from that 
through the optomagnonic Barnett effect
of the opposite circular polarization $\eta =\pm $,
spins in the left (right) AF are aligned along the +(-) $z$ axis,
see Fig.~\ref{fig:junction} (cf., Table~\ref{tab:Compare}).
Consequently, 
the direction of the macroscopic coherent spin precession 
in the left BEC phase
becomes opposite to the one in the right.
Therefore
the relative phase becomes the sum of the frequency
as
$ \theta (t)
=( \Omega _{\mathrm{L}}+\Omega _{\mathrm{R}})t+\theta (0)$
with
$ {\textrm{sgn}}(\Omega _{\mathrm{L}})
= {\textrm{sgn}}(\Omega _{\mathrm{R}})$.
On the other hand, 
since spins in the FM-FM junction of Ref.~\cite{KKPD}
are aligned along the same direction,
the relative phase is characterized 
essentially as the difference of the frequency
$\omega _{\mathrm{L(R)}}$
of the left (right) BEC as
$ (\omega _{\mathrm{L}}-\omega _{\mathrm{R}})t $
where
$ {\textrm{sgn}}(\omega _{\mathrm{L}})
= {\textrm{sgn}}(\omega _{\mathrm{R}})$.
For the details of the frequency
$\omega _{\mathrm{L(R)}}$,
see Ref.~\cite{KKPD}.
Thus,
using the scheme of Fig.~\ref{fig:junction}
we can enhance the frequency of the Josephson oscillation.
For all of these reasons, 
ultrafast spin transport
is realized in our AF-AF junction.
Those are summarized in Tables~\ref{tab:CompareJosephson0}
and~\ref{tab:CompareJosephson}.

Several comments on our approach are in order.
First, 
in this paper we assume 
not the easy-axis anisotropy 
but
the easy-plane anisotropy.
Therefore a spin-flop transition~\cite{SpinFlopTra} is absent in this setup~\cite{Supple}.
Second,
we find that a dc spin Josephson effect 
might be induced
but realized unstably
in this setup~\cite{Supple}.
Third,
for the difference between
the inverse Faraday effect~\cite{Kirilyuk,KimelNatureIFaraday}
and
the optical Barnett effect~\cite{OtherOpticalBarnett2,OtherOpticalBarnett3},
i.e., the laser-induced magnetization~\cite{FloquetST2,FloquetST},
see Ref.~\cite{KNST_OBarnett}.
Last,
throughout this paper,
we have assumed a sufficiently low temperature
where phonon degrees of freedom
ceases to work~\cite{magnon10mK,tabuchi,tabuchiScience,Tmagnonphonon}.
It will be interesting to study the effect of phonons
on the spin Josephson effect,
which we leave for future work.

We remark that Ref.~\cite{spinSFscienceAdvanced}
reported experimental signatures of spin superfluid 
in ${\mathrm{Cr}}_2{\mathrm{O}}_3$
subjected to a strong magnetic field along the easy-axis.
For the generation of the spin superfluid~\cite{bunkov},
the easy-plane anisotropy is essential,
while originally 
${\mathrm{Cr}}_2{\mathrm{O}}_3$
possesses the easy-axis anisotropy.
From this,
it is expected that 
the applied magnetic field changes the spin anisotropy of
${\mathrm{Cr}}_2{\mathrm{O}}_3$
and effectively makes it the easy-plane.
Thus we can control the spin anisotropy.
Still, 
to find the insulating AF
which intrinsically possesses the perfect easy-plane anisotropy,
i.e., the $U(1)$ spin-rotational symmetry
within the easy-plane,
is of significance.
To the best of our knowledge,
this remains a challenge of the antiferromagnetic spintronics study.

\section{Conclusion}
\label{sec:Sec.11}
Using the macroscopic coherence of 
the optical magnon Bose-Einstein condensates
intrinsic to insulating antiferromagnets,
we have proposed the optomagnonic Josephson effect.
The optomagnonic Barnett effect 
associated with the fully spin-polarized state
enables us to exploit coherent magnons of high frequency
over the conventional ones
of (sub-) terahertz associated with the N\'eel magnetic order.
Applying the optomagnonic Barnett effect 
to the junction of weakly coupled two insulating antiferromagnets,
we have shown that the ultrafast spin Josephson effect  
of those optical magnon Bose-Einstein condensates
is realized.
The period of the optomagnonic Josephson oscillation 
is much shorter 
than the conventional one of the order of picoseconds.
Our work builds a bridge between optics and magnonics,
and is expected to become the key ingredient for
the ultrafast manipulation of spin information.

\acknowledgements

The author would like to sincerely thank S. Takayoshi
for fruitful discussions,
especially for sharing his technical note;
a part of the Appendices is based on it.
The author is grateful also to
K. A. van Hoogdalem, P. Simon, and D. Loss
for the collaborative work on the related topic,
Y. Korai for helpful discussions at the early stage of this work,
S. K. Kim for useful information and comments on this manuscript,
and M. Oka for helpful feedback on this work.
The author is supported 
by JSPS KAKENHI Grant Number JP20K14420
and 
by Leading Initiative for Excellent Young Researchers, MEXT, Japan.

\appendix

\renewcommand\thefigure{\thesection.\arabic{figure}}

\section{Classical spin theory}
\label{sec:classical}
In this Appendix we derive the critical frequency $\Omega_{\mathrm{c}}$
for the fully spin-polarized state of AFs,
and evaluate the magnetization along the $z$ axis as a function of laser frequency (Fig.~\ref{fig:Magnetization}).
First, we consider the antiferromagnetic model 
described by the Hamiltonian
\begin{align}
 \hat{\mathcal{H}}_{0}
   =J\sum_{\langle i,j \rangle}
     \hat{\boldsymbol{S}}_{i}\cdot
     \hat{\boldsymbol{S}}_{j}
     +D\sum_{i}(\hat{S}_{i}^{z})^{2}.
\label{eq:Hamil_AF_SM}
\end{align}
The easy-plane single ion anisotropy $D>0$ 
stabilizes the N\'eel magnetic order on the $xy$ plane. 
Under the application of circularly polarized laser, 
the effective Hamiltonian
reduces to 
(see the main text)
\begin{align}
 {\hat{\mathcal{H}}}_{\mathrm{eff}}
   =&\hat{\mathcal{H}}_{0}
     -\eta\hbar\Omega\sum_{i}\hat{S}_{i}^{z}.
\label{eq:Hamil_eff_SM}
\end{align}
The AF consists of the sublattice A and B.
The classical spin configuration is determined 
in the way that the energy per spin $\epsilon$,
\begin{subequations}
\begin{align}
 \epsilon=&\frac{E}{N}  \\
   =&\frac{z_{0}J}{2}\boldsymbol{S}_{\mathrm{A}}\cdot\boldsymbol{S}_{\mathrm{B}}
     +\frac{D}{2}[(S_{\mathrm{A}}^{z})^{2}+(S_{\mathrm{B}}^{z})^{2}]
     -\frac{\eta\hbar\Omega}{2}(S_{\mathrm{A}}^{z}+S_{\mathrm{B}}^{z})
\label{eq:Classical_E_SM}
\end{align}
\end{subequations}
is minimized, 
where 
$N$ is the number of spin sites, 
$E$ denotes the total energy, 
$z_{0}$ represents the coordination number,
and
$\boldsymbol{S}_{\mathrm{A(B)}}$
is the spin on the sublattice A (B).

\setcounter{figure}{0}
\begin{figure}[t]
\begin{center}
\includegraphics[width=8.5cm,clip]{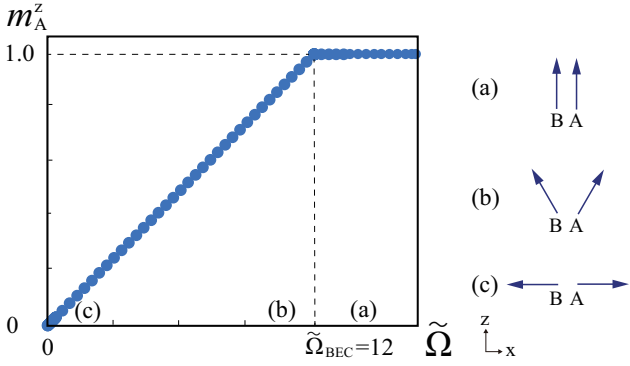}
\caption{Plot of the magnetization along the $z$ axis 
$ m_{\mathrm{A}}^{z} = m_{\mathrm{B}}^{z}  $
for $\eta =1$
as a function of 
the rescaled frequency
$\tilde{\Omega }:=\hbar \Omega /J $ 
obtained by numerically solving Eq.~\eqref{eq:energy},
i.e., minimizing the energy $\epsilon$,
with the experimental parameter values given in the main text.
The optical Barnett field develops the magnetization continuously.
The absence of the first order transition,
i.e., jump of $m_{\mathrm{A(B)}}^{z} $,
assures the validity of the description 
in terms of magnons. 
(a)
In the high frequency regime 
$ \tilde{\Omega } > \tilde{\Omega }  _{\mathrm{BEC}}=12 $,
spins are fully polarized along the $z$ axis.
(b)
Decreasing the frequency,
the optical magnon BEC transition is provoked on the point
$ \tilde{\Omega } = \tilde{\Omega }  _{\mathrm{BEC}} $,
where the AF acquires a transverse component of 
local magnetization
associated with the spontaneous $U(1)$ symmetry breaking.
Thus a macroscopic coherent state is formed.
(c)
In the low frequency regime $ \tilde{\Omega }\sim 0$,
the N\'eel magnetic order is developed on the $xy$ plane
and we take it the $x$ axis without loss of generality.
Throughout this paper we study the optical magnon BEC 
(b) in the vicinity of 
$ \tilde{\Omega } = \tilde{\Omega }  _{\mathrm{BEC}} $.
}
\label{fig:Magnetization}
\end{center}
\end{figure}

Next, we focus on the vicinity of 
the critical frequency $\Omega_{\mathrm{c}}$ (Fig.~\ref{fig:Magnetization}). 
Since Eq.~\eqref{eq:Classical_E_SM} has the ${\mathrm{U}}(1)$ symmetry,
without loss of generality we assume that 
$\boldsymbol{S}_{\mathrm{A}}$
and
$\boldsymbol{S}_{\mathrm{B}}$
are in the $xz$ plane as
$\boldsymbol{S}_{\mathrm{A}}=( S\sin\theta, 0, \eta S\cos\theta)=:(m_{\mathrm{A}}^{x}, 0, m_{\mathrm{A}}^{z})$ and 
$\boldsymbol{S}_{\mathrm{B}}=(-S\sin\theta, 0, \eta S\cos\theta)=:(m_{\mathrm{B}}^{x}, 0, m_{\mathrm{B}}^{z})$.
Then Eq.~\eqref{eq:Classical_E_SM} is rewritten as 
\begin{subequations}
\begin{align}
 \epsilon
   =&\frac{z_{0}JS^{2}}{2}\cos(2\theta)
     +DS^{2}\cos^{2}\theta
     -\hbar\Omega S\cos\theta  \\
   =&(z_{0}J+D)S^{2}\cos^{2}\theta
     -\hbar\Omega S\cos\theta-\frac{z_{0}JS^{2}}{2}.
\label{eq:energy}
\end{align}
\end{subequations}
We call the $0<\theta<\pi$ case as the V-shape phase, 
which corresponds to the magnon BEC phase 
as we see below, 
since the sublattice magnetization $\boldsymbol{S}_{\mathrm{A}}$ and 
$\boldsymbol{S}_{\mathrm{B}}$ 
form the V-shape. 
For the stability of the V-shape phase $0<\theta<\pi$, 
the condition 
\begin{align}
 z_{0}J+D>0
\label{eq:condition1}
\end{align}
is necessary. 
This condition corresponds to the repulsive interaction 
between magnons in the spin wave theory (cf., Appendix~\ref{sec:spinwave}). 
The energy of Eq.~\eqref{eq:Classical_E_SM} 
takes the minimum 
at $\theta=0$ 
for $\hbar\Omega\geq 2(z_{0}J+D)S$ 
and at 
$\theta=\arccos \{{\hbar\Omega}/[{2(z_{0}J+D)S}]\} \neq 0$ 
for $\hbar\Omega<2(z_{0}J+D)S$.
Thus the critical frequency 
for the fully spin-polarized state of AFs
is given as 
\begin{align}
 \hbar\Omega_{\mathrm{c}}=2(z_{0}J+D)S.
\label{eq:critfreq}
\end{align}

Finally, 
magnetization along the $z$ axis per spin is given as 
\begin{subequations}
\begin{align}
 \eta S^{z}=&S\cos\theta  \\
   =&\frac{\hbar\Omega}{2(z_{0}J+D)}  \\
   =&\frac{\Omega}{\Omega_{\mathrm{c}}}S.
\label{eq:mz}
\end{align}
\end{subequations}
The numerical plot is provided in Fig.~\ref{fig:Magnetization}.

\section{Magnon theory}
\label{sec:spinwave}
In this Appendix we derive the transition point
for the magnon BEC,
${\Omega}_{\mathrm{BEC}}$,
associated with the fully spin-polarized state in the high frequency regime,
and evaluate the number of magnon condensates
in the vicinity of $\Omega = {\Omega}_{\mathrm{BEC}}$.
The ground state is fully polarized 
$\boldsymbol{S}=(0,0,\eta S)$ 
for $\Omega>\Omega_{\mathrm{c}}$. 
First,
we perform the Holstein-Primakoff transformation,
\begin{align}
 \eta \hat{S}_{i}^{z}
   &=S-\hat{n}_{i}, \nonumber  \\
 \hat{S}_{i}^{x}+\eta i\hat{S}_{i}^{y}
   &=\sqrt{2S}\Big(1-\frac{\hat{n}_{i}}{2S}\Big)^{1/2}\hat{a}_{i},  \nonumber  \\
 \hat{S}_{i}^{x}-\eta i\hat{S}_{i}^{y}
   &=\sqrt{2S}\hat{a}_{i}^{\dagger}\Big(1-\frac{\hat{n}_{i}}{2S}\Big)^{1/2},
\nonumber
\end{align}
where $\hat{a}_{i}^{\dagger}$ and $\hat{a}_{i}$ are 
creation and annihilation operators for bosons, i.e., magnons, 
and $\hat{n}_{i}\equiv \hat{a}_{i}^{\dagger}\hat{a}_{i}$ 
is the number operator. 
We make an expansion and 
retain up to the fourth order terms of $\hat{a}_{i}$ and $\hat{a}_{i}^{\dagger}$, 
\begin{align}
 \eta \hat{S}_{i}^{z}
 &=S-\hat{n}_{i}, \nonumber  \\
 \hat{S}_{i}^{x}+\eta i\hat{S}_{i}^{y}
   &=\sqrt{2S}\Big(1-\frac{\hat{n}_{i}}{4S}\Big)\hat{a}_{i}, \nonumber  \\
 \hat{S}_{i}^{x}-\eta i\hat{S}_{i}^{y}
   &=\sqrt{2S}\hat{a}_{i}^{\dagger}\Big(1-\frac{\hat{n}_{i}}{4S}\Big).
\nonumber
\end{align}
Using the magnon operator, 
the Hamiltonian [Eq.~\eqref{eq:Hamil_eff_SM}] 
is rewritten as 
\begin{align}
 {\hat{\mathcal{H}}}_{\mathrm{eff}}
   &=JS\sum_{\langle i,j \rangle}
     (\hat{a}_{i}^{\dagger}\hat{a}_{j}+\mathrm{H.c.})
   -\frac{J}{4}\sum_{\langle i,j \rangle}
     (\hat{a}_{i}^{\dagger}\hat{n}_{i}\hat{a}_{j}
     +\hat{a}_{i}^{\dagger}\hat{n}_{j}\hat{a}_{j}+\mathrm{H.c.}) \nonumber  \\
   &-z_{0}JS\sum_{i}\hat{n}_{i}
   +J\sum_{\langle i,j \rangle}\hat{n}_{i}\hat{n}_{j} 
   -2DS\sum_{i}\hat{n}_{i} \nonumber  \\
   &+D\sum_{i}\hat{n}_{i}^{2}
   +\hbar\Omega\sum_{i}\hat{n}_{i},
\end{align}
where constant terms are dropped. 
We consider the cubic lattice 
and the configuration number is $z_{0}=6$. 
After the Fourier transform 
for the positional vector $\boldsymbol{r}_{i}$ as
\begin{align}
 \hat{a}_{\boldsymbol{k}}
   &=\sqrt{\frac{1}{N}}\sum_{i}
     e^{-i\boldsymbol{k}\cdot\boldsymbol{r}_{i}}
     \hat{a}_{i}, \nonumber  \\
 \hat{a}_{\boldsymbol{k}}^{\dagger}
   &=\sqrt{\frac{1}{N}}\sum_{i}
     e^{i\boldsymbol{k}\cdot\boldsymbol{r}_{i}}
     \hat{a}_{i}^{\dagger}, \nonumber  \\
 \hat{n}_{\boldsymbol{k}}
   &=\hat{a}_{\boldsymbol{k}}^{\dagger}
     \hat{a}_{\boldsymbol{k}},\nonumber
\end{align} 
we obtain 
\begin{align}
 {\hat{\mathcal{H}}}_{\mathrm{eff}}
   &=2JS\sum_{\boldsymbol{k}}
     [\cos(k_{x}a)+\cos(k_{y}a)+\cos(k_{z}a)]
     \hat{n}_{\boldsymbol{k}} \nonumber  \\
   &+(-z_{0}JS-2DS+\hbar\Omega)
     \sum_{\boldsymbol{k}}\hat{n}_{\boldsymbol{k}}
   +\hat{\mathcal{U}},
\label{eq:HamilMagnon4th}
\end{align}
where $a$ is the lattice constant. 
The interaction term $\hat{\mathcal{U}}$ is represented as 
\begin{widetext}
\begin{align}
 \hat{\mathcal{U}}=&-\frac{J}{2N}
     \sum_{\boldsymbol{k}_{1},\boldsymbol{k}_{2},
     \boldsymbol{k}_{3},\boldsymbol{k}_{4}}
     [\cos(k_{1,x}a)+\cos(k_{1,y}a)+\cos(k_{1,z}a)]
     \hat{a}_{\boldsymbol{k}_{1}}^{\dagger}\hat{a}_{\boldsymbol{k}_{2}}^{\dagger}
     \hat{a}_{\boldsymbol{k}_{3}}\hat{a}_{\boldsymbol{k}_{4}}
     \delta_{\boldsymbol{k}_{1}+\boldsymbol{k}_{2},
             \boldsymbol{k}_{3}+\boldsymbol{k}_{4}}
\nonumber\\
   &-\frac{J}{2N}
     \sum_{\boldsymbol{k}_{1},\boldsymbol{k}_{2},
     \boldsymbol{k}_{3},\boldsymbol{k}_{4}}
     [\cos(k_{4,x}a)+\cos(k_{4,y}a)+\cos(k_{4,z}a)]
     \hat{a}_{\boldsymbol{k}_{1}}^{\dagger}\hat{a}_{\boldsymbol{k}_{2}}^{\dagger}
     \hat{a}_{\boldsymbol{k}_{3}}\hat{a}_{\boldsymbol{k}_{4}}
     \delta_{\boldsymbol{k}_{1}+\boldsymbol{k}_{2},
             \boldsymbol{k}_{3}+\boldsymbol{k}_{4}}
\nonumber\\
   &+\frac{J}{N}
     \sum_{\boldsymbol{k}_{1},\boldsymbol{k}_{2},
     \boldsymbol{k}_{3},\boldsymbol{k}_{4}}
     [\cos((k_{1,x}-k_{2,x})a)+\cos((k_{1,y}-k_{2,y})a)+\cos((k_{1,z}-k_{2,z})a)]
     \hat{a}_{\boldsymbol{k}_{1}}^{\dagger}\hat{a}_{\boldsymbol{k}_{2}}
     \hat{a}_{\boldsymbol{k}_{3}}^{\dagger}\hat{a}_{\boldsymbol{k}_{4}}
     \delta_{\boldsymbol{k}_{1}+\boldsymbol{k}_{3},
             \boldsymbol{k}_{2}+\boldsymbol{k}_{4}}
\nonumber\\
   &+\frac{D}{N}
     \sum_{\boldsymbol{k}_{1},\boldsymbol{k}_{2},
     \boldsymbol{k}_{3},\boldsymbol{k}_{4}}
     \hat{a}_{\boldsymbol{k}_{1}}^{\dagger}\hat{a}_{\boldsymbol{k}_{2}}
     \hat{a}_{\boldsymbol{k}_{3}}^{\dagger}\hat{a}_{\boldsymbol{k}_{4}}
     \delta_{\boldsymbol{k}_{1}+\boldsymbol{k}_{3},
             \boldsymbol{k}_{2}+\boldsymbol{k}_{4}}.
\label{eq:interaction}
\end{align}
\end{widetext}
The magnon Hamiltonian in the corotating frame [Eq.~\eqref{eq:HamilMagnon4th}]
consists of
the kinetic energy,
the magnon-magnon interaction,
and
the Zeeman energy of the magnetic field in the magnet.
The effective magnetic field 
${\mathbf{B}} _{\mathrm{eff}}=(B _{\mathrm{eff}}^x, 0, B _{\mathrm{eff}}^z) $
magnons acquire in the corotating frame is
$B _{\mathrm{eff}}^z:= \hbar (\Omega -\Omega _{\mathrm{BEC}})+6JS$
along the $z$ axis,
while
$B _{\mathrm{eff}}^x:= B_0$
along the $x$ axis.
Since
$6JS=O(10){\mathrm{meV}} \sim O(10^2){\mathrm{T}} $
and consequently
$6JS  \gg B_0  $ in general,
the effective magnetic field along the $x$ axis
$B _{\mathrm{eff}}^x= B_0$
is negligibly small
compared with the $z$ component
$B _{\mathrm{eff}}^x \ll B _{\mathrm{eff}}^z  $
even in the vicinity of 
$\Omega \approx \Omega _{\mathrm{BEC}} $.
When $\Omega$ is decreased from the large value, 
the band, 
$2JS[\cos(k_{x}a)+\cos(k_{y}a)+\cos(k_{z}a)]-z_{0}JS-2DS+\hbar\Omega$, 
touches the zero energy at the wavenumber 
$\boldsymbol{k}=\boldsymbol{\pi}\coloneqq (\pi/a,\pi/a,\pi/a)$. 
Therefore the magnons created by 
$\hat{a}_{\boldsymbol{\pi}}^{\dagger}$ condensate at
\begin{align}
 \hbar\Omega_{\mathrm{BEC}}=2(6J+D)S,
\label{eq:BECfreq}
\end{align}
which coincides with
$\hbar\Omega_{\mathrm{c}}$ 
[Eq.~\eqref{eq:critfreq}].

Next, 
we consider the interaction term. 
Since magnons condensate at $\boldsymbol{k}=\boldsymbol{\pi}$, 
we only keep the term with 
$\boldsymbol{k}_{1}=\boldsymbol{k}_{2}=\boldsymbol{k}_{3}=\boldsymbol{k}_{4}=\boldsymbol{\pi}$ 
in Eq.~\eqref{eq:interaction} as
\begin{subequations}
 \begin{align}
{\hat{\mathcal{U}}}=&
\frac{3J}{N}
   \hat{a}_{\boldsymbol{\pi}}^{\dagger}
   \hat{a}_{\boldsymbol{\pi}}^{\dagger}
   \hat{a}_{\boldsymbol{\pi}}
   \hat{a}_{\boldsymbol{\pi}}
 +\frac{3J+D}{N}
   \hat{a}_{\boldsymbol{\pi}}^{\dagger}
   \hat{a}_{\boldsymbol{\pi}}
   \hat{a}_{\boldsymbol{\pi}}^{\dagger}
   \hat{a}_{\boldsymbol{\pi}}  \\
 =&\frac{6J+D}{N}
{\hat{a}}^{\dagger }_{\boldsymbol{\pi}}{\hat{a}}^{\dagger }_{\boldsymbol{\pi}}{\hat{a}}_{\boldsymbol{\pi}}{\hat{a}}_{\boldsymbol{\pi}}
+\frac{3J+D}{N} {\hat{n}}_{\boldsymbol{\pi}}.
\end{align}
\end{subequations}
Thus $6J+D>0$ corresponds to 
repulsive interaction. 
The $\boldsymbol{k}=\boldsymbol{\pi}$ sector 
in the Hamiltonian of 
Eq.~\eqref{eq:HamilMagnon4th} is given as 
 \begin{align}
 {\hat{{\mathcal{H}}}}_{\mathrm{eff}}(\boldsymbol{k}=\boldsymbol{\pi})
 &=(-12JS-2DS+\hbar \Omega +\frac{3J+D}{N}) {\hat{n}}_{\boldsymbol{\pi}}
 \nonumber  \\
 &+\frac{6J+D}{N}
{\hat{a}}^{\dagger }_{\boldsymbol{\pi}}{\hat{a}}^{\dagger }_{\boldsymbol{\pi}}{\hat{a}}_{\boldsymbol{\pi}}{\hat{a}}_{\boldsymbol{\pi}}.
\end{align}
Since we treat a macroscopic system, the number of spin sites $N$ is large enough to approximate as 
\begin{subequations}
 \begin{align}
 {\hat{{\mathcal{H}}}}_{\mathrm{eff}}(\boldsymbol{k}=\boldsymbol{\pi})
 & \simeq    (-12JS-2DS+\hbar \Omega) {\hat{n}}_{\boldsymbol{\pi}} \nonumber  \\
 &+\frac{6J+D}{N}
{\hat{a}}^{\dagger }_{\boldsymbol{\pi}}{\hat{a}}^{\dagger }_{\boldsymbol{\pi}}{\hat{a}}_{\boldsymbol{\pi}}{\hat{a}}_{\boldsymbol{\pi}}
  \label{eq:HamilPi}  \\
&= \hbar (\Omega -\Omega_{\mathrm{BEC}}){\hat{n}}_{\boldsymbol{\pi}}
+\frac{\hbar \Omega_{\mathrm{BEC}}}{2NS}
{\hat{a}}^{\dagger }_{\boldsymbol{\pi}}{\hat{a}}^{\dagger }_{\boldsymbol{\pi}}{\hat{a}}_{\boldsymbol{\pi}}{\hat{a}}_{\boldsymbol{\pi}}.
\label{eq:HamilPi2}
\end{align}
\end{subequations}
In order for the magnon BEC state 
with finite $\langle \hat{n}_{\boldsymbol{\pi}}\rangle$ 
to be stabilized, 
the repulsive interaction $6J+D>0$ is necessary, 
which corresponds to Eq.~\eqref{eq:condition1}.

Finally,
by minimizing Eq.~\eqref{eq:HamilPi}
we obtain 
\begin{subequations}
\begin{align}
 \langle\hat{n}_{\boldsymbol{\pi}}\rangle
   =&\frac{12JS+2DS-\hbar\Omega}{2(6J+D)}N \\
   =&\frac{\Omega_{\mathrm{BEC}}-\Omega}{\Omega_{\mathrm{BEC}}}NS.
\label{eq:nvalue}
\end{align}
\end{subequations}
Thus the number of magnon condensates
is characterized as a function of laser frequency $\Omega $ 
for $\Omega <\Omega_{\mathrm{BEC}}$.
The magnetization along the $z$ axis per spin is provided as 
\begin{subequations}
\begin{align}
 \frac{\langle\eta\sum_{i}\hat{S}_{i}^{z}\rangle}{N}
   =&S-\frac{\langle\sum_{i}\hat{n}_{i}\rangle}{N} \\
 \simeq &S-\frac{\langle\hat{n}_{\boldsymbol{\pi}}\rangle}{N} \\
   =&\frac{\Omega}{\Omega_{\mathrm{BEC}}}S,
\end{align}
\end{subequations}
which corresponds to Eq.~\eqref{eq:mz}.

\section{Reference frame}
\label{sec:reference}
In this Appendix we give the description in the original stationary reference frame. 
Note that in Appendices~\ref{sec:classical} and \ref{sec:spinwave}
we describe the system in the rotating frame.
We represent the transformation,
$R:=\exp(\eta i\Omega tS^{z})$,
to the rotating frame. 
The observables transform as 
\begin{align}
 (\tilde{S}^{x},\tilde{S}^{y})
 =&R^{-1}(S^{x},S^{y})R  \nonumber \\
   =&(\cos(\Omega t)S^{x}+\eta\sin(\Omega t)S^{y},
     -\eta\sin(\Omega t)S^{x}+\cos(\Omega t)S^{y}).
\nonumber
\end{align}
The Heisenberg equation of motion is as 
\begin{align}
 i\partial_{t}\tilde{\mathcal{O}}
   =&i\partial_{t}(R^{\dagger}\mathcal{O}R) \nonumber \\
   =&(i\partial_{t}R^{\dagger})\mathcal{O}R
     +R^{\dagger}(i\partial_{t}\mathcal{O})R
     +R^{\dagger}\mathcal{O}(i\partial_{t}R)
\nonumber\\
   =&(i\partial_{t}R^{\dagger})R\tilde{\mathcal{O}}
     +R^{\dagger}[\mathcal{O},\mathcal{H}]R
     +\tilde{\mathcal{O}}R^{\dagger}(i\partial_{t}R) \nonumber \\
   =&[\tilde{\mathcal{O}},
     \tilde{\mathcal{H}}+R^{\dagger}(i\partial_{t}R)].
\nonumber
\end{align}
Thus the effective Hamiltonian is given as
$\tilde{\mathcal{H}}+R^{\dagger}(i\partial_{t}R)
=\tilde{\mathcal{H}}-\eta\Omega S^{z}$.

The purpose of this Appendix is to
give the description 
in the original stationary reference frame. 
First,
we represent the magnon operators in the reference frame 
as $\hat{b}^{(\dagger)}$, i.e., 
\begin{align}
 \hat{a}_{\boldsymbol{k}}^{(\dagger)}
   =\hat{R}^{\dagger}\hat{b}_{\boldsymbol{k}}^{(\dagger)}\hat{R},
\end{align}
where 
$\hat{R}=\exp(\eta i\Omega t\hat{S}_{\mathrm{tot}}^{z})
=\exp[i\Omega t(NS-\sum_{\boldsymbol{k}}
\hat{a}_{\boldsymbol{k}}^{\dagger}\hat{a}_{\boldsymbol{k}})]$.
The time-evolution is evaluated as
\begin{widetext}
\begin{align}
 i\hbar\partial_{t}\hat{b}_{\boldsymbol{k}}^{(\dagger)}
   =&i\hbar\partial_{t}(\hat{R}\hat{a}_{\boldsymbol{k}}^{(\dagger)}\hat{R}^{\dagger}) \nonumber \\
   =&(i\hbar\partial_{t}\hat{R})\hat{a}_{\boldsymbol{k}}^{(\dagger)}\hat{R}^{\dagger}
     +\hat{R}(i\hbar\partial_{t}\hat{a}_{\boldsymbol{k}}^{(\dagger)})\hat{R}^{\dagger}
     +\hat{R}\hat{a}_{\boldsymbol{k}}^{(\dagger)}(i\hbar\partial_{t}\hat{R}^{\dagger})
\nonumber\\
   =&-\hbar\Omega \hat{R}(NS-\sum_{\boldsymbol{k}}
     \hat{a}_{\boldsymbol{k}}^{\dagger}\hat{a}_{\boldsymbol{k}})
     \hat{R}^{\dagger}
     \hat{R}\hat{a}_{\boldsymbol{k}}^{(\dagger)}\hat{R}^{\dagger}
     +\hat{R}[\hat{a}_{\boldsymbol{k}}^{(\dagger)},\hat{\mathcal{H}}]\hat{R}^{\dagger}
     +\hbar\Omega \hat{R}\hat{a}_{\boldsymbol{k}}^{(\dagger)}\hat{R}^{\dagger}
     \hat{R}(NS-\sum_{\boldsymbol{k}}
     \hat{a}_{\boldsymbol{k}}^{\dagger}\hat{a}_{\boldsymbol{k}})
     \hat{R}^{\dagger}
\nonumber\\
   =&-\hbar\Omega (NS-\sum_{\boldsymbol{k}}
     \hat{b}_{\boldsymbol{k}}^{\dagger}\hat{b}_{\boldsymbol{k}})
     \hat{b}_{\boldsymbol{k}}^{(\dagger)}
     +[\hat{b}_{\boldsymbol{k}}^{(\dagger)},\hat{R}\hat{\mathcal{H}}\hat{R}^{\dagger}]
     +\hbar\Omega \hat{b}_{\boldsymbol{k}}^{(\dagger)}
     (NS-\sum_{\boldsymbol{k}}
     \hat{b}_{\boldsymbol{k}}^{\dagger}\hat{b}_{\boldsymbol{k}})
\nonumber\\
   =&[\hat{b}_{\boldsymbol{k}}^{(\dagger)},
     \hat{R}\hat{\mathcal{H}}\hat{R}^{\dagger}
     +\hbar\Omega(NS-\sum_{\boldsymbol{k}}
     \hat{b}_{\boldsymbol{k}}^{\dagger}\hat{b}_{\boldsymbol{k}})]
\nonumber\\
   =&[\hat{b}_{\boldsymbol{k}}^{(\dagger)},
     \hat{R}\hat{\mathcal{H}}\hat{R}^{\dagger}
     -\hbar\Omega\sum_{\boldsymbol{k}}
     \hat{b}_{\boldsymbol{k}}^{\dagger}\hat{b}_{\boldsymbol{k}}].
\label{eq:EffHamiltonianBECaaa}
\end{align}
\end{widetext}
Next,
we focus on the $\boldsymbol{k}=\boldsymbol{\pi}$ sector. 
The effective Hamiltonian in the rotating frame
is given as Eq.~\eqref{eq:HamilPi2};
\begin{align}
 {\hat{\mathcal{H}}}_{\mathrm{eff}}(\boldsymbol{k}=\boldsymbol{\pi})
   =\hbar(\Omega -\Omega_{\mathrm{BEC}})
     \hat{a}_{\boldsymbol{\pi}}^{\dagger}\hat{a}_{\boldsymbol{\pi}}
     +\frac{\hbar\Omega_{\mathrm{BEC}}}{2NS}
     \hat{a}_{\boldsymbol{\pi}}^{\dagger}\hat{a}_{\boldsymbol{\pi}}^{\dagger}
     \hat{a}_{\boldsymbol{\pi}}\hat{a}_{\boldsymbol{\pi}}.
\nonumber
\end{align}
Finally,
from Eq.~\eqref{eq:EffHamiltonianBECaaa}
we obtain 
the corresponding Hamiltonian
in the original stationary reference frame as
\begin{subequations}
\begin{align}
 \hat{R}{\hat{\mathcal{H}}}_{\mathrm{eff}}(\boldsymbol{k}=\boldsymbol{\pi})
   \hat{R}^{\dagger}
     -\hbar\Omega
     \hat{b}_{\boldsymbol{\pi}}^{\dagger}\hat{b}_{\boldsymbol{\pi}}
   &=-\hbar\Omega_{\mathrm{BEC}}
     \hat{b}_{\boldsymbol{\pi}}^{\dagger}\hat{b}_{\boldsymbol{\pi}} \nonumber \\
     &+\frac{\hbar\Omega_{\mathrm{BEC}}}{2NS}
     \hat{b}_{\boldsymbol{\pi}}^{\dagger}\hat{b}_{\boldsymbol{\pi}}^{\dagger}
     \hat{b}_{\boldsymbol{\pi}}\hat{b}_{\boldsymbol{\pi}}  \\
   &=:{\hat{\mathcal{H}}}_{\boldsymbol{k}=\boldsymbol{\pi}}.
\end{align}
\end{subequations}
The equation of motion is given as 
\begin{subequations}
\begin{align}
 i\hbar\partial_{t}
   \hat{b}_{\boldsymbol{\pi}}
   =&[\hat{b}_{\boldsymbol{\pi}},
     -\hbar\Omega_{\mathrm{BEC}}
     \hat{b}_{\boldsymbol{\pi}}^{\dagger}\hat{b}_{\boldsymbol{\pi}}
     +\frac{\hbar\Omega_{\mathrm{BEC}}}{2NS}
     \hat{b}_{\boldsymbol{\pi}}^{\dagger}\hat{b}_{\boldsymbol{\pi}}^{\dagger}
     \hat{b}_{\boldsymbol{\pi}}\hat{b}_{\boldsymbol{\pi}}]
\nonumber\\
   =&-\hbar\Omega_{\mathrm{BEC}}
     \hat{b}_{\boldsymbol{\pi}}
     +\frac{\hbar\Omega_{\mathrm{BEC}}}{NS}
     \hat{b}_{\boldsymbol{\pi}}^{\dagger}
     \hat{b}_{\boldsymbol{\pi}}\hat{b}_{\boldsymbol{\pi}}, \\
 i\hbar\partial_{t}
   \hat{b}_{\boldsymbol{\pi}}^{\dagger}
   =&[\hat{b}_{\boldsymbol{\pi}}^{\dagger},
     -\hbar\Omega_{\mathrm{BEC}}
     \hat{b}_{\boldsymbol{\pi}}^{\dagger}\hat{b}_{\boldsymbol{\pi}}
     +\frac{\hbar\Omega_{\mathrm{BEC}}}{2NS}
     \hat{b}_{\boldsymbol{\pi}}^{\dagger}\hat{b}_{\boldsymbol{\pi}}^{\dagger}
     \hat{b}_{\boldsymbol{\pi}}\hat{b}_{\boldsymbol{\pi}}]
\nonumber\\
   =&\hbar\Omega_{\mathrm{BEC}}
     \hat{b}_{\boldsymbol{\pi}}^{\dagger}
     -\frac{\hbar\Omega_{\mathrm{BEC}}}{NS}
     \hat{b}_{\boldsymbol{\pi}}^{\dagger}
     \hat{b}_{\boldsymbol{\pi}}^{\dagger}\hat{b}_{\boldsymbol{\pi}}.
\end{align}
\end{subequations}
If we approximate 
$\hat{b}_{\boldsymbol{\pi}}^{\dagger}\hat{b}_{\boldsymbol{\pi}}
=\hat{a}_{\boldsymbol{\pi}}^{\dagger}\hat{a}_{\boldsymbol{\pi}}
\simeq
[({\Omega_{\mathrm{BEC}}-\Omega})/{\Omega_{\mathrm{BEC}}}]
NS$, 
cf., Eq.~\eqref{eq:nvalue},
those reduce to
\begin{align}
 i\hbar\partial_{t}
   \hat{b}_{\boldsymbol{\pi}}
   =-\hbar\Omega
     \hat{b}_{\boldsymbol{\pi}},\quad
 i\hbar\partial_{t}
   \hat{b}_{\boldsymbol{\pi}}^{\dagger}
   =\hbar\Omega
     \hat{b}_{\boldsymbol{\pi}}^{\dagger}.
\end{align}
These equations represent the precession 
with the frequency $\Omega$ 
in synchronization with the laser field.

\section{Optomagnonic Josephson equation}
In this Appendix starting from the Hamiltonian 
${\hat{\mathcal{H}}}_{\mathrm{tot}}={\hat{\mathcal{H}}}_{\mathrm{L}}+{\hat{\mathcal{H}}}_{\mathrm{R}}+{\hat{V}}$
for the junction 
of weakly coupled two magnon BECs 
(see the main text),
we derive the optomagnonic Josephson equations in the main text.
First,
the Heisenberg equation of motion provides
\begin{subequations}
\begin{align}
i\hbar \frac{d{{{\hat{b}}}}_{\mathrm{L}}}{dt}
    =& [{\hat{b}}_{\mathrm{L}}, {\hat{\mathcal{H}}}_{\mathrm{L}}+{\hat{V}}]  \\
    =& \hbar \Omega_{\mathrm{L}}  {{\hat{b}}}_{\mathrm{L}} 
      + 2 U_{\mathrm{L}} {{\hat{b}}}^{\dagger }_{\mathrm{L}} {{\hat{b}}}_{\mathrm{L}}  {{\hat{b}}}_{\mathrm{L}} 
       - K {{\hat{b}}}^{\dagger }_{\mathrm{R}}, \\
i\hbar \frac{d{{{\hat{b}}}}_{\mathrm{R}}}{dt}
    =& [{\hat{b}}_{\mathrm{R}}, {\hat{\mathcal{H}}}_{\mathrm{R}}+{\hat{V}}]  \\
    =& \hbar \Omega_{\mathrm{R}}  {{\hat{b}}}_{\mathrm{R}} 
      + 2 U_{\mathrm{R}} {{\hat{b}}}^{\dagger }_{\mathrm{R}} {{\hat{b}}}_{\mathrm{R}}  {{\hat{b}}}_{\mathrm{R}} 
       - K {{\hat{b}}}^{\dagger }_{\mathrm{L}}.
\label{eq:two-state2SM}
\end{align}
\end{subequations}
Taking the expectation value $\langle {\hat{b}}_{\mathrm{L(R)}} \rangle =:b_{\mathrm{L(R)}} \in \mathbb{C}$, 
we obtain the two-state model in the main text.

Next,
noting that
$(d b_{\mathrm{L(R)}}/dt)/b_{\mathrm{L(R)}} = (d/dt){\mathrm{ln}}b_{\mathrm{L(R)}}$
and
multiplying the two-state model by $1/b_{\mathrm{L(R)}}$,
it is recast into
\begin{subequations}
\begin{align}
i\hbar \frac{d}{dt} {\mathrm{ln}}b_{\mathrm{L}}
    =& \hbar \Omega_{\mathrm{L}}  
      + 2 U_{\mathrm{L}} N_{\mathrm{L}}  
       - K \frac{{{b}}^{\dagger }_{\mathrm{R}}}{{b_{\mathrm{L}}}},
\label{eq:two-state5SM}  \\
i\hbar \frac{d}{dt} {\mathrm{ln}}b_{\mathrm{R}}
    =& \hbar \Omega_{\mathrm{R}}   
      + 2 U_{\mathrm{R}} N_{\mathrm{R}} 
       - K \frac{{{b}}^{\dagger }_{\mathrm{L}}}{b_{\mathrm{R}}}.
\label{eq:two-state6SM}
\end{align}
\end{subequations}
Since $b_{\mathrm{L}}(t) =\sqrt{N_{\mathrm{L}}(t)} {\mathrm{e}}^{i\theta _{\mathrm{L}}(t)} $
and $b_{\mathrm{R}}(t) =\sqrt{N_{\mathrm{R}}(t)} {\mathrm{e}}^{-i\theta _{\mathrm{R}}(t)}$,
those are rewritten as
\begin{subequations}
\begin{align}
i\hbar \Big( \frac{1}{2}\frac{1}{N_{\mathrm{L}}}\frac{dN_{\mathrm{L}}}{dt}+i\frac{d\theta _{\mathrm{L}}}{dt} \Big)
    &= \hbar \Omega_{\mathrm{L}}  
      + 2 U_{\mathrm{L}} N_{\mathrm{L}}  \nonumber \\
       &- K \sqrt{\frac{N_{\mathrm{R}}}{N_{\mathrm{L}}}}{\mathrm{e}}^{i(\theta _{\mathrm{R}}-\theta _{\mathrm{L}})},
\label{eq:two-state7SM}  \\
i\hbar \Big( \frac{1}{2}\frac{1}{N_{\mathrm{R}}}\frac{dN_{\mathrm{R}}}{dt}-i\frac{d\theta _{\mathrm{R}}}{dt} \Big)
    &= \hbar \Omega_{\mathrm{R}}  
      + 2 U_{\mathrm{R}} N_{\mathrm{R}}  \nonumber \\
       &- K \sqrt{\frac{N_{\mathrm{L}}}{N_{\mathrm{R}}}}{\mathrm{e}}^{i(\theta _{\mathrm{R}}-\theta _{\mathrm{L}})}.
\label{eq:two-state8SM}
\end{align}
\end{subequations}
Dividing Eq.~\eqref{eq:two-state7SM} into 
the real and imaginary parts, we obtain
\begin{subequations}
\begin{align}
-\hbar \frac{d\theta _{\mathrm{L}}}{dt}
    =& (\hbar \Omega _{\mathrm{L}}+2 U_{\mathrm{L}} N_{\mathrm{L}}) 
    - K\sqrt{\frac{N_{\mathrm{R}}}{{N_{\mathrm{L}}}}} {\mathrm{cos}}(\theta _{\mathrm{R}} - \theta _{\mathrm{L}}),
\label{eq:two-state9SM} \\
\hbar  \frac{dN _{\mathrm{L}}}{dt}
    =&-2K \sqrt{N_{\mathrm{L}}N_{\mathrm{R}}}{\mathrm{sin}}(\theta _{\mathrm{R}} - \theta _{\mathrm{L}}).
\label{eq:two-state10SM}
\end{align}
\end{subequations}
In the same way, 
Eq.~\eqref{eq:two-state8SM} provides
\begin{subequations}
\begin{align}
\hbar \frac{d\theta _{\mathrm{R}}}{dt}
    =& (\hbar \Omega _{\mathrm{R}}+2 U_{\mathrm{R}} N_{\mathrm{R}}) 
    - K\sqrt{\frac{N_{\mathrm{L}}}{{N_{\mathrm{R}}}}} {\mathrm{cos}}(\theta _{\mathrm{R}} - \theta _{\mathrm{L}}), 
\label{eq:two-state11SM} \\
\hbar  \frac{dN _{\mathrm{R}}}{dt}
    =&-2K \sqrt{N_{\mathrm{L}}N_{\mathrm{R}}}{\mathrm{sin}}(\theta _{\mathrm{R}} - \theta _{\mathrm{L}}).
\label{eq:two-state12SM}
\end{align}
\end{subequations}

Here, we remark that
the calculation of
[Eq.~\eqref{eq:two-state10SM}]$-$[Eq.~\eqref{eq:two-state12SM}] gives
\begin{align}
\frac{d}{dt}(N_{\mathrm{L}}-N_{\mathrm{R}})
   =0.  
\label{eq:N_{-}SM}
\end{align}
This means that the total spin angular momentum is conserved
and $N_{-}:=N_{\mathrm{L}}-N_{\mathrm{R}}$ is constant.
On the other hand, 
the calculation of
[Eq.~\eqref{eq:two-state10SM}]$+$[Eq.~\eqref{eq:two-state12SM}] provides
\begin{align}
\frac{d}{dt}(N _{\mathrm{L}}+N _{\mathrm{R}})
   =-\frac{4K}{\hbar}\sqrt{N_{\mathrm{L}}N_{\mathrm{R}}}{\mathrm{sin}}(\theta _{\mathrm{R}} - \theta _{\mathrm{L}}).
\label{eq:N_{+}SM}
\end{align}
This describes the magnonic Josephson spin current flowing across the junction interface.
Introducing $N_{+}(t):=N_{\mathrm{L}}(t)+N_{\mathrm{R}}(t) >0$
and defining $ z(t):={N_{+}(t)}/{N_{-}}$, it satisfies
\begin{align}
 \mid z(t)\mid \geq 1.
\label{eq:ztheta}
\end{align}
In this work, without loss of generality 
we assume the initial condition $N_{-}(0)>0$ for convenience. 
Since $N_{-}:=N_{\mathrm{L}}-N_{\mathrm{R}}$ is constant,
this ensures $z(t)\geq 1$
and
\begin{subequations}
\begin{align}
z^2=&\frac{N_-^2+4N_{\mathrm{L}} N_{\mathrm{R}}}{N_-^2}  \\
      =&1+4\frac{N_{\mathrm{L}} N_{\mathrm{R}}}{N_-^2},
\label{eq:z2}
\end{align}
\end{subequations}
resulting in
\begin{align}
\frac{\sqrt{N_{\mathrm{L}}N_{\mathrm{R}}}}{N_-}=\frac{\sqrt{z^2-1}}{2}.
\label{eq:z2-2}
\end{align}

Finally,
using the relation,
from Eq.~\eqref{eq:N_{+}SM} 
we obtain 
\begin{align}
\frac{dz(t)}{dt}
    = -\frac{2K}{\hbar}\sqrt{z(t)^2-1}{\mathrm{sin}}\theta (t),
\label{eq:JosephsonEqzSM}
\end{align}
where $\theta (t) := \theta _{\mathrm{R}}(t)-\theta _{\mathrm{L}}(t)$
is the relative phase.
The calculation of
[Eq.~\eqref{eq:two-state9SM}] $+$ [Eq.~\eqref{eq:two-state11SM}] gives
\begin{align}
\hbar \frac{d}{dt}(\theta _{\mathrm{R}}-\theta _{\mathrm{L}})
    &= (\hbar \Omega _{\mathrm{L}}+\hbar \Omega _{\mathrm{R}})
    +2 (U_{\mathrm{L}} N_{\mathrm{L}}+U_{\mathrm{R}} N_{\mathrm{R}})   \nonumber \\
    &- K\Big( \sqrt{\frac{N_{\mathrm{R}}}{{N_{\mathrm{L}}}}}+\sqrt{\frac{N_{\mathrm{L}}}{{N_{\mathrm{R}}}}} \Big) 
   {\mathrm{cos}}(\theta _{\mathrm{R}} - \theta _{\mathrm{L}}).
\label{eq:two-stateThetaSM}
\end{align}
Since 
\begin{subequations}
\begin{align}
\sqrt{\frac{N_{\mathrm{R}}}{{N_{\mathrm{L}}}}}+\sqrt{\frac{N_{\mathrm{L}}}{{N_{\mathrm{R}}}}}
=&\frac{2}{\sqrt{z^2-1}}z,  \\
 U_{\mathrm{L}} N_{\mathrm{L}}+U_{\mathrm{R}} N_{\mathrm{R}}
=&\frac{U_{\mathrm{L}}+U_{\mathrm{R}}}{2}N_{-} z
  +\frac{U_{\mathrm{L}}-U_{\mathrm{R}}}{2}N_{-},
\label{eq:tech}
\end{align}
\end{subequations}
Eq.~\eqref{eq:two-stateThetaSM} is rewritten as
\begin{align}
\frac{d\theta (t)}{dt}
    &= \Big[(\Omega _{\mathrm{L}}+\Omega _{\mathrm{R}})
      +\frac{U _{\mathrm{L}}-U _{\mathrm{R}}}{\hbar}N_{-} \Big] 
    +\Big(\frac{U _{\mathrm{L}}+U _{\mathrm{R}}}{\hbar}N_{-} \Big) z(t)   \nonumber \\
    &- \frac{2K}{\hbar}\frac{z(t)}{\sqrt{z(t)^2-1}}{\mathrm{cos}}\theta (t).
\label{eq:JosephsonEqThetaSM}
\end{align}
Eqs.~\eqref{eq:JosephsonEqzSM} 
and~\eqref{eq:JosephsonEqThetaSM} 
are the optomagnonic Josephson equation in the main text.

We remark that 
introducing the normalized time 
\begin{align}
\tau :=\frac{2K}{\hbar}t,
\label{eq:tau}
\end{align}
the optomagnonic Josephson equations 
[Eqs.~\eqref{eq:JosephsonEqzSM} 
and~\eqref{eq:JosephsonEqThetaSM}] 
are recast into the dimensionless form as 
\begin{subequations}
\begin{align}
\frac{dz(\tau)}{d\tau}
    &= - \sqrt{z(\tau)^2-1}{\mathrm{sin}}\theta (\tau), 
\label{eq:JosephsonEqtauz} \\
\frac{d\theta (\tau)}{d\tau}
    &= \Big[\frac{\hbar(\Omega _{\mathrm{L}}+\Omega _{\mathrm{R}})}{2K}
      +\frac{U _{\mathrm{L}}-U _{\mathrm{R}}}{2K}N_{-} \Big] \nonumber \\
     &+\Big(\frac{U _{\mathrm{L}}+U _{\mathrm{R}}}{2K}N_{-} \Big) z(\tau) 
     - \frac{z(\tau)}{\sqrt{z(\tau)^2-1}}{\mathrm{cos}}\theta (\tau). 
\label{eq:JosephsonEqtauTheta}
\end{align}
\end{subequations}

\section{Tunneling amplitude}
In this Appendix we estimate the tunneling amplitude in spin language.
Due to a finite overlap of the wave functions of the localized spins that reside on 
the relevant two-dimensional boundaries of each insulator,
there exists in general a finite exchange interaction between the boundary spins.
Let us assume that it is described 
by the boundary spin Hamiltonian as
${\hat{V}}_{\mathrm{s}}=
-J_{\mathrm{tun}} \hat{\boldsymbol{S}}_{\mathrm{L}} \cdot \hat{\boldsymbol{S}}_{\mathrm{R}} $,
where
$\hat{\boldsymbol{S}}_{\mathrm{L(R)}}$
is the spin operator for the boundary spins
forming the macroscopic coherent state;
the spin quantum number 
in the left (right) insulator 
is
$S_{\mathrm{L(R)}}$
and 
$J_{\mathrm{tun}}$
of 
$|J_{\mathrm{tun}}| \ll J$
is the weak spin exchange interaction between the boundary spins.
By means of the magnon theory,
it reduces to the tunneling Hamiltonian ${\hat{V}}$ in the main text
as
${\hat{V}}_{\mathrm{s}}\approx  
-J_{\mathrm{tun}}
\sqrt{S_{\mathrm{L}}S_{\mathrm{R}}}
( {\hat{b}}_{\mathrm{L}} {\hat{b}}_{\mathrm{R}}+{\hat{b}}^{\dagger }_{\mathrm{L}}{\hat{b}}^{\dagger }_{\mathrm{R}})$.
Thus we find that the tunneling amplitude is represented 
in spin language as 
\begin{align}
\mid K\mid =\mid J_{\mathrm{tun}}\mid \sqrt{S_{\mathrm{L}}S_{\mathrm{R}}}.
\end{align}
Note that ${\mathrm{sgn}}(K)={\mathrm{sgn}}(J_{\mathrm{tun}})=\pm $ in general, 
see the main text.

\section{An analysis on optomagnonic dc Josephson effect}
In this Appendix under the assumption that
magnon BECs are realized stably,
we discuss an attempt to realize 
an optomagnonic dc Josephson effect.
Assuming the initial condition $z(0) \gg 1$
and
tuning the parameters as
$\hbar (\Omega _{\mathrm{L}}+\Omega _{\mathrm{R}})
   +({U _{\mathrm{L}}-U _{\mathrm{R}}})N_{-}=0$
and ${U _{\mathrm{L}}+U _{\mathrm{R}}}=0$,
the optomagnonic Josephson equation in the main text
is recast into
\begin{subequations}
\begin{align}
\frac{dz(\tau)}{d\tau}\mid _{\tau\ll 1}
    =& -z(\tau){\mathrm{sin}}\theta (\tau),
\label{eq:JosephsonEqztau} \\
\frac{d\theta (\tau)}{d\tau}\mid _{\tau\ll 1}
    =& -{\mathrm{cos}}\theta (\tau),
\label{eq:JosephsonEqtau} 
\end{align}
\end{subequations}
where $\tau:=(2K/\hbar)t$ is the normalized time.
Noting that ${d\theta (\tau)}/({d\tau})\mid _{\tau\ll 1}=0$ 
when $\theta (0)=\pm \pi/2 $,
we find that
the functions,
$z(\tau)\mid _{\tau\ll 1}= -z(0)\tau +z(0) $
and $z(\tau)\mid _{\tau\ll 1}= z(0)\tau +z(0) $,
approximately satisfy the Josephson equation
for $\theta (0)=\pi/2$ and $\theta (0)=-\pi/2$, respectively.
This implies that
the dc Josephson effect
satisfying
${dz (\tau)}/({d\tau})\mid _{\tau\ll 1}=(\textrm{const.})$
and
${d\theta (\tau)}/({d\tau})\mid _{\tau\ll 1}=0$
is induced for $\tau \ll  1$.

From this, one might suspect that
the dc Josephson effect is realizable.
However,
it requires the condition ${U _{\mathrm{L}}+U _{\mathrm{R}}}=0$, 
which means that the magnon-magnon interaction is attractive in one side.
Therefore, the magnon BEC state itself is unstable in one side 
as long as one employs this setup.

\bibliography{PumpingRef}

\end{document}